\documentclass{mem}
\usepackage[sort&compress,numbers]{natbib}
\usepackage{txfonts}
\usepackage{balance}
\usepackage{graphicx}
\usepackage[a4paper,breaklinks,dvipdfm]{hyperref}

\idline{75}{282}

\begin{document}


\newcommand{\vla}{VLA}
\newcommand{\most}{MOST}
\newcommand{\park}{Parkes}
\newcommand{\atc}{ATCA}
\newcommand{\rosat}{{\it ROSAT}}
\newcommand{\rxt}{{\it RXTE}}
\newcommand{\xmm}{{\it XMM-Newton}}
\newcommand{\chan}{{\it Chandra}}
\newcommand{\sax}{{\it BeppoSAX}}
\newcommand{\asca}{{\it ASCA}}
\newcommand{\suz}{{\it Suzaku}}
\newcommand{\intg}{{\it INTEGRAL}}
\newcommand{\ib}{IBIS}
\newcommand{\spi}{SPI}
\newcommand{\isg}{ISGRI}
\newcommand{\ibisg}{IBIS/ISGRI}
\newcommand{\swi}{{\it SWIFT}}
\newcommand{\bat}{BAT}
\newcommand{\gro}{{\it CGRO}}
\newcommand{\cpt}{COMPTEL}
\newcommand{\egr}{EGRET}
\newcommand{\agile}{{\it AGILE}}
\newcommand{\fermi}{{\it Fermi}}
\newcommand{\hess}{H.E.S.S.}

\newcommand{\rxj}{RX~J1713.7$-$3946}
\newcommand{\velajr}{Vela~Jr}
\newcommand{\rcw}{RCW~86}
\newcommand{\sn}{SN~1006}
\newcommand{\hessj}{HESS~J1731$-$347}

\newcommand{\un}[1]{~\hspace{-1pt}\ensuremath{\mathrm{#1}}}
\def\edot{$\dot{{\rm E}}$}

\def\ks{km s$^{-1}$}
\def\kms{$\mathrm {km s}^{-1}$}
\def\d{$^\circ$}
\def\m{$^\prime$}
\def\s{$^{\prime\prime}$}
\def\hh{$^{\mathrm h}$}
\def\mm{$^{\mathrm m}$}
\def\second{$^{\mathrm s}$}
\def\cm3{cm$^{-3}$}
\def\pp{$^{\prime\prime}$}
\def\msun{M$_\odot$}

\def\eg{{\it e.g.~}}
\def\etal{et~al.~}
\def\ie{{\em i.e.~}}

\title{
Supernova Remnants and Pulsar Wind Nebulae in the 
Cherenkov Telescope Array era
}

   \subtitle{}

\author{
M. \,Renaud\inst{1} for the CTA consortium
          }

  \offprints{M. Renaud}

\institute{
Laboratoire Univers et Particules de Montpellier, 
Universit\'e Montpellier 2, CNRS/IN2P3, CC 072, 
Place Eug\`ene Bataillon, F-34095 Montpellier Cedex 5, France
\email{mrenaud@lupm.univ-montp2.fr}
}

\authorrunning{M. Renaud}

\titlerunning{SNRs and PWNe in the CTA era}

\abstract{The Cherenkov Telescope Array (CTA) is planned to serve as a 
ground-based observatory for (very-)high-energy gamma-ray astronomy, open to a 
wide astrophysics community, providing a deep insight into the non-thermal 
high-energy universe. It foresees a factor of $\sim$10 improvement in 
sensitivity above 100 GeV, with substantially better angular and spectral 
resolutions and wider field-of-view in comparison with currently operational 
experiments. The CTA consortium is investigating the different physics cases 
for different proposed array configurations and subsets. Pulsar Wind Nebulae
(PWNe), the most numerous VHE Galactic sources, and Supernova Remnants (SNRs), 
believed to be the acceleration sites of the bulk of cosmic rays, will be two 
of the main observation targets for CTA. In this contribution, the main 
scientific goals regarding PWNe and SNRs are discussed, and quantitative 
examples of the capability of CTA to achieve these objectives are presented.
\keywords{Gamma rays: general -- Surveys -- 
ISM: supernova remnants, cosmic rays}
}
\maketitle{}

\section{VHE astronomy and the CTA observatory}
\label{s:intro}

While only a handful of sources were known to emit in the very-high-energy 
(VHE; E $>$ 100 GeV) gamma-ray domain until 2005, more than 120 sources have 
now been detected (two third lying in the Galaxy)\footnote{See 
\texttt{http://tevcat.uchicago.edu/}}, thanks to the current 
generation of Imaging Atmospheric Telescopes (IACTs) such as \hess, VERITAS 
and MAGIC. The end-products of massive star evolution, Supernova Remnants 
(SNRs) and Pulsar Wind Nebulae (PWNe), are thought to be the main sites of 
particle acceleration, and are characterized by non-thermal emission over the 
entire electromagnetic spectrum through several mechanisms (synchrotron [SC], 
inverse-Compton [IC], non-thermal bremsstrahlung, proton-proton interactions 
and subsequent $\pi^0$~decay). The \hess~experiment, in particular, has imaged 
the shell-type morphology of several young ($\lesssim$ a few kyr) SNRs and has 
revealed a new population of middle-aged ($\gtrsim$ 10 kyr) VHE-emitting PWNe 
\citep{c:hinton09}. Although these observations demonstrate that SNRs and PWNe 
are able to accelerate particles in the $\sim$10--100 TeV range, recent 
theoretical investigations have raised several issues on particle acceleration 
at (non-)relativistic shocks, within the general paradigm of the nature of the 
cosmic-ray (CR) sources in the Galaxy \citep[see \eg][]{c:blasi10}.

\begin{figure}[htb!]
\centering
\includegraphics[width=0.48\textwidth]{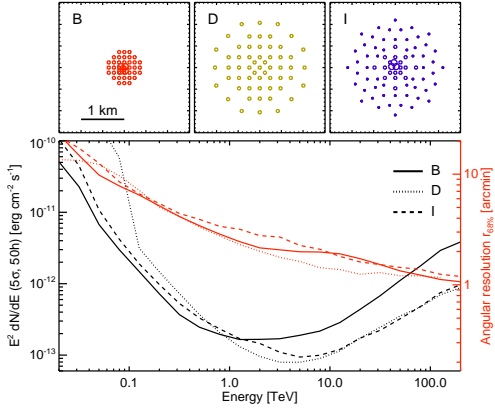}
\caption{\footnotesize {\it Upper panels:} Three representative CTA array 
         layouts dubbed B, D and I. Telescope sizes are not drawn to scale. 
         {\it Lower panel:} Point-source sensitivity (5 $\sigma$ in 50 h, in 
         black) and angular resolution (68\% containment radius, in red) of 
         the three configurations. B is optimized in the low-energy ($<$ 1 TeV)
         domain, D provides the best sensitivity at energies $>$ 1 TeV, and I 
         is optimized for providing better performance over the whole energy 
         range.
}
\label{f:fig1}
\end{figure}

The Cherenkov Telescope Array \citep{c:cta10}, currently under its 
Preparatory Phase (2011--2013), is an initiative to build an advanced facility 
for ground-based HE/VHE (20 GeV--200 TeV) gamma-ray astronomy, with various 
scientific themes, from cosmic energetic particles (origin of CRs, jets and 
shocks) to fundamental physics (Dark Matter, Lorenz Invariance violation). The 
consortium is firmly embedded in the European processes guiding fields of 
astronomy and astroparticle physics. It gathers $\gtrsim$800 scientists and 
engineers in $\gtrsim$100 institutes. The construction of the full array of 
50--100 $\sim$23--24 m-, $\sim$10--12 m- and $\sim$5--8 m-diameter Cherenkov 
telescopes is planned to be completed around 2018. The main characteristics 
of three representative CTA configurations, currently considered within the 
consortium, are presented in Figure \ref{f:fig1}. A point-source sensitivity
at the level of one mCrab and an angular resolution of 1--2 arcmin at TeV 
energies are foreseen with such layouts. In the following, Monte-Carlo 
simulations of shell-type SNRs and PWNe as seen with CTA are presented, with 
an emphasis on spectro-imaging ($\S$ \ref{s:specima}) and population ($\S$ 
\ref{s:popu}) studies. The results presented here are based on simulations 
performed for observations at a zenith angle of 20\d, assuming a symmetrical 
gaussian point spread function (PSF)  with a 68\% containment radius as given 
in Figure \ref{f:fig1}, while neglecting the effects of the finite energy 
resolution of the instrument (which are mitigated for the large energy ranges 
considered). 

\section{Spectro-imaging studies} 
\label{s:specima}

\begin{figure*}[!htb]
\centering
\includegraphics[width=0.497\textwidth]{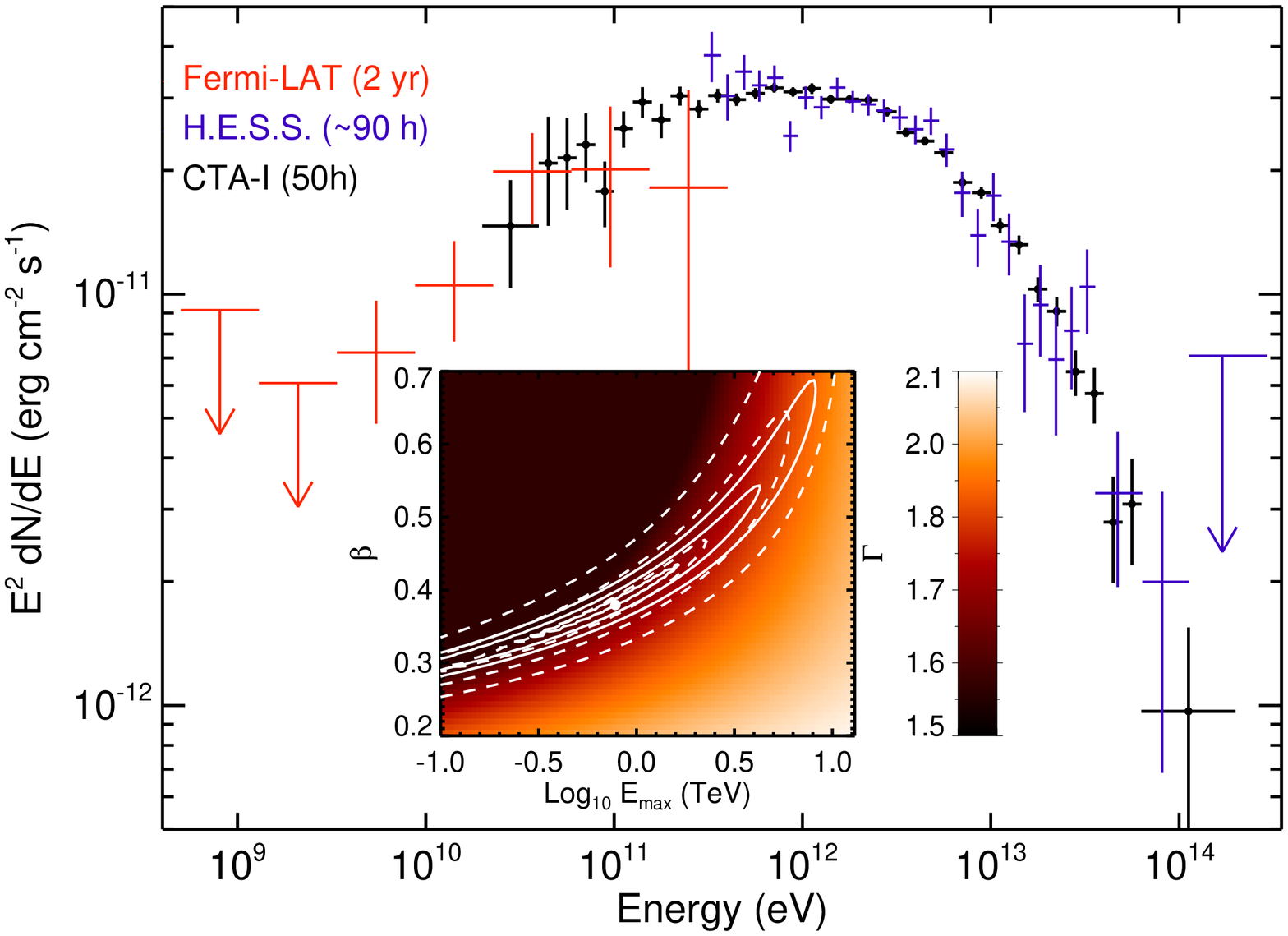}
\includegraphics[width=0.497\textwidth]{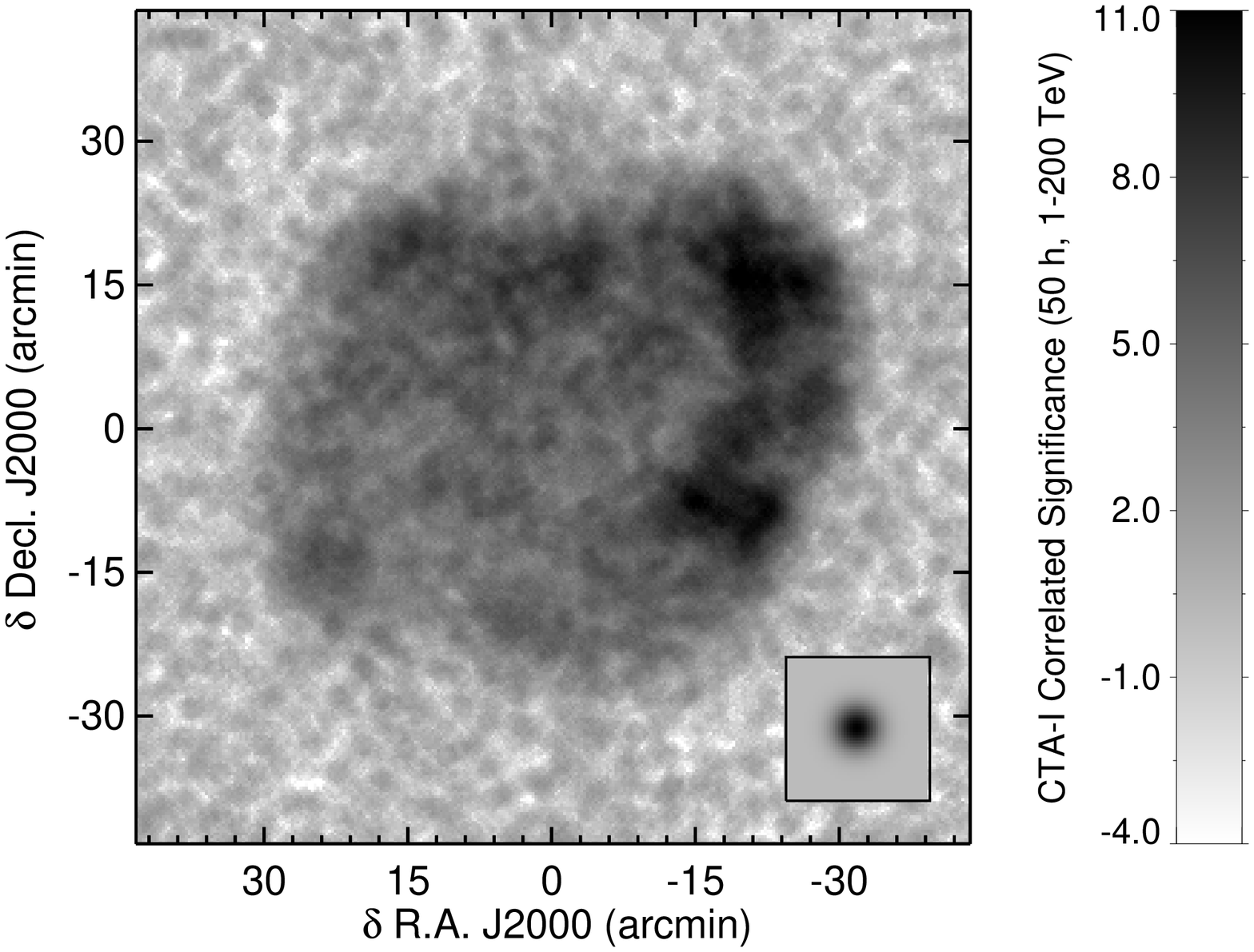}
\caption{\footnotesize CTA-I simulations of \rxj~for 50 h of observing time.
         {\it Left:} Spectral energy distribution (SED, in black) with the 
         best-fitted parameters of the joint \fermi/LAT (in red) and \hess~(in 
         blue) spectrum. The inset figure shows the 1, 2, and 3 $\sigma$ 
         confidence ellipses in the E$_{{\rm cut}}$-$\beta$ plane, derived 
         from the fit of 100 simulated spectra (solid lines) and of the 
         \fermi/LAT-\hess~spectrum (dashed lines), superimposed on the image 
         of the $\Gamma$ values. The white dot gives the best-fit values: 
         $\Gamma$ $\sim$1.62, E$_{{\rm cut}}$ $\sim$0.78 TeV and $\beta$ 
         $\sim$0.38. {\it Right:} Correlated significance map 
         (R$_{{\rm corr}}$ = 0.02\d) above 1 TeV. The \xmm~image of 
         \rxj~\citep{c:acero09} has been used as template. The CTA-I PSF is 
         shown in the inset.
}
\label{f:fig2}
\end{figure*}

\rxj~is the brightest shell-type SNR in the VHE domain, originally detected 
with \hess~\citep{c:aharonian07}, and more recently with \fermi/LAT in the HE 
domain \citep{c:abdo11}. The nature of the gamma-ray emission (IC emission 
from leptons vs. $\pi^0$ decay from hadrons) is still debated, even though IC 
emission seems to be favored from a spectral point of view. The joint 
\fermi/LAT and \hess~spectrum was fit with an exponential cutoff power-law in 
the form N$_0$ E$^{-\Gamma}$ exp(-(E/E$_{{\rm cut}}$)$^{{\rm \beta}}$). The 
best-fitted parameters have been used to simulate CTA spectra of \rxj~\citep[whose morphology is assumed to be that measured with \hess,][]{c:aharonian06}, 
as shown in Figure \ref{f:fig2} (left), for CTA-I and 50 h of observing time. 
Besides reducing greatly the uncertainties on each spectral parameter, CTA 
will be able to probe the highest energy domain. As shown by 
\citet{c:morlino09}, and later confirmed by \citet{c:ellison10}, the predicted 
IC emission from accelerated leptons, while accounting for the radio and X-ray 
measurements of the associated SC emission, could not explain the flux 
measured with \hess~above $\sim$35 TeV (though with low S/N ratios of 2.5, 1.5 
and 0.6). Therefore, any significant flux detected at the highest energies 
would point towards the existence of $\gtrsim$ 200 TeV hadrons still confined 
in the shell. For this purpose, 10$^3$ simulations of the \rxj~spectrum above 
35 TeV in 50 h have been carried out. CTA-I/D are clearly favored over B, with 
mean S/N ratios of $\sim$7 and 6, respectively.

Figure \ref{f:fig2} (right) shows a CTA-I simulated image of \rxj~above 1 TeV
in 50 h, assuming as the morphological and spectral templates the \xmm~image 
\citep{c:acero09} and the best-fit on the joint \fermi/LAT-\hess~SED, 
respectively. Spatial structures such as the double shell-like morphology in 
the W-NW region are clearly seen. Simulated spectra have also been extracted 
over 0.25\d-wide boxes, as defined in \citet{c:aharonian06}. The mean errors 
on $\Gamma$ and E$_{{\rm cut}}$ (with $\beta$ fixed) are estimated to be of 
the order of 0.08 and 0.3 TeV (90\% confidence level), respectively, \ie~small 
enough to allow the exploration of possible azimuthal variations of these 
parameters, in combination with what is performed nowadays in X-rays on the 
SC spectrum of many young SNRs.

\section{Population studies}
\label{s:popu}

On one hand, the small number of detected TeV shells (RX~J0852$-$4622 aka 
\velajr, \rxj, \rcw, \sn~and \hessj) prevents one from performing population 
studies in order to shed further light on particle acceleration in these 
sources, even when considering the other (unresolved) detected young shell-type
SNRs (such as Cas~A and Tycho). On the other hand, $\sim$30 VHE PWNe have been 
detected so far, mainly in the inner regions of the Galaxy surveyed with \hess,
most of them being extended on scales of $\sigma$ $\sim$ 0.2\d. Moreover, a 
significant fraction of the unassociated VHE sources could be such ancient 
nebulae, as suggested by \citet{c:dejager09}. This scenario seems to be 
conforted by the on-going detection of many gamma-ray pulsars with \fermi/LAT, 
several of which lying close to VHE sources
\citep [\eg MGRO~J1908$+$06/PSR~J1907$+$0602,][]{c:abdo10}. It is then worth 
estimating the expected number of shell-like SNRs and PWNe to be detected with 
CTA, in order to (1) trigger SNR/PWN population studies and (2) quantify the 
level of source confusion along the Galactic Plane (and especially in the 
regions close to the spiral arm tangents), given the large number of 
already-detected VHE PWNe and PWN candidates. In the following, simulations 
have been carried out assuming a reasonable exposure time of 20~h in the survey
of the Galactic Plane. 

\begin{figure}[htb!]
\centering
\includegraphics[width=0.239\textwidth]{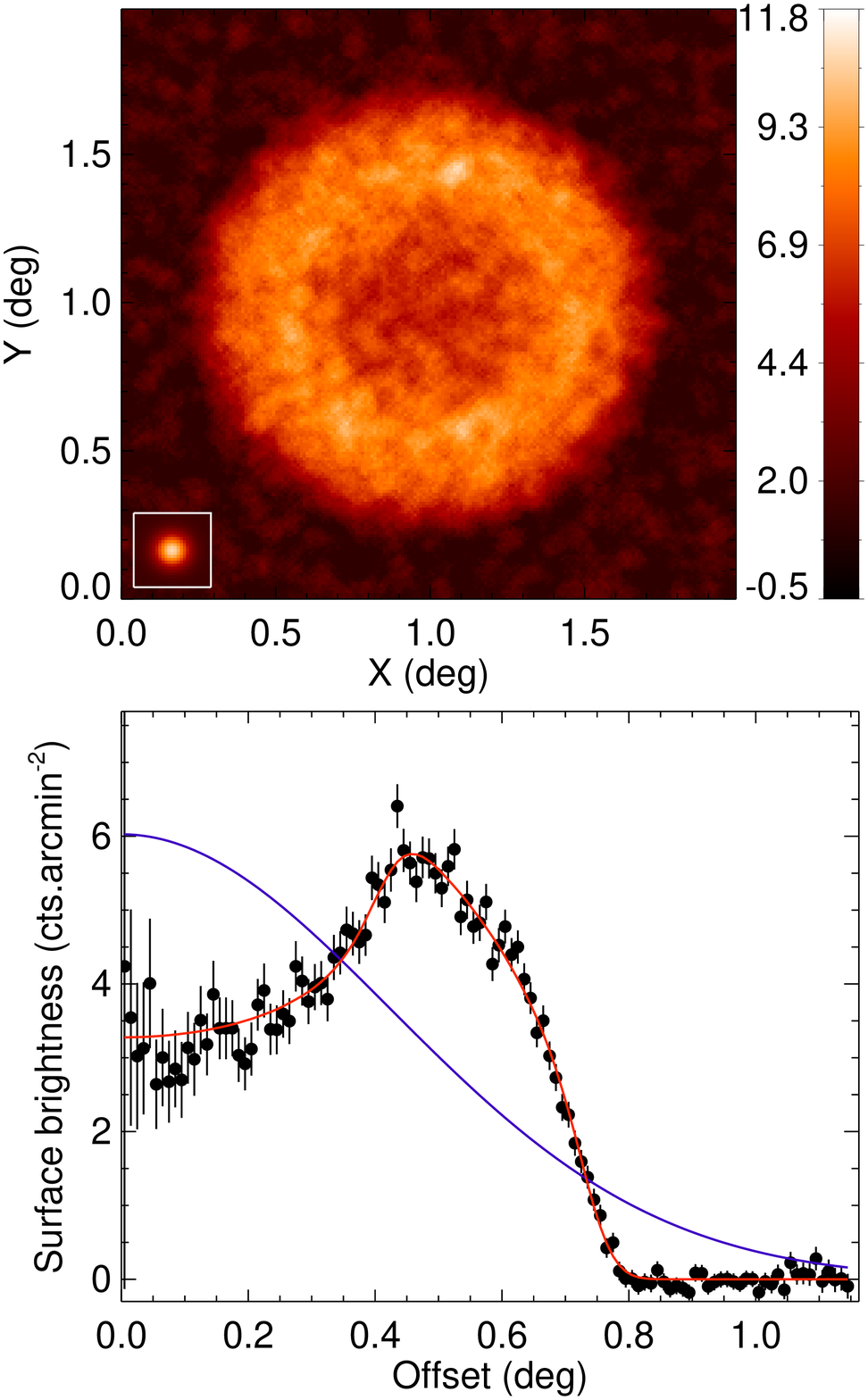}
\includegraphics[width=0.239\textwidth]{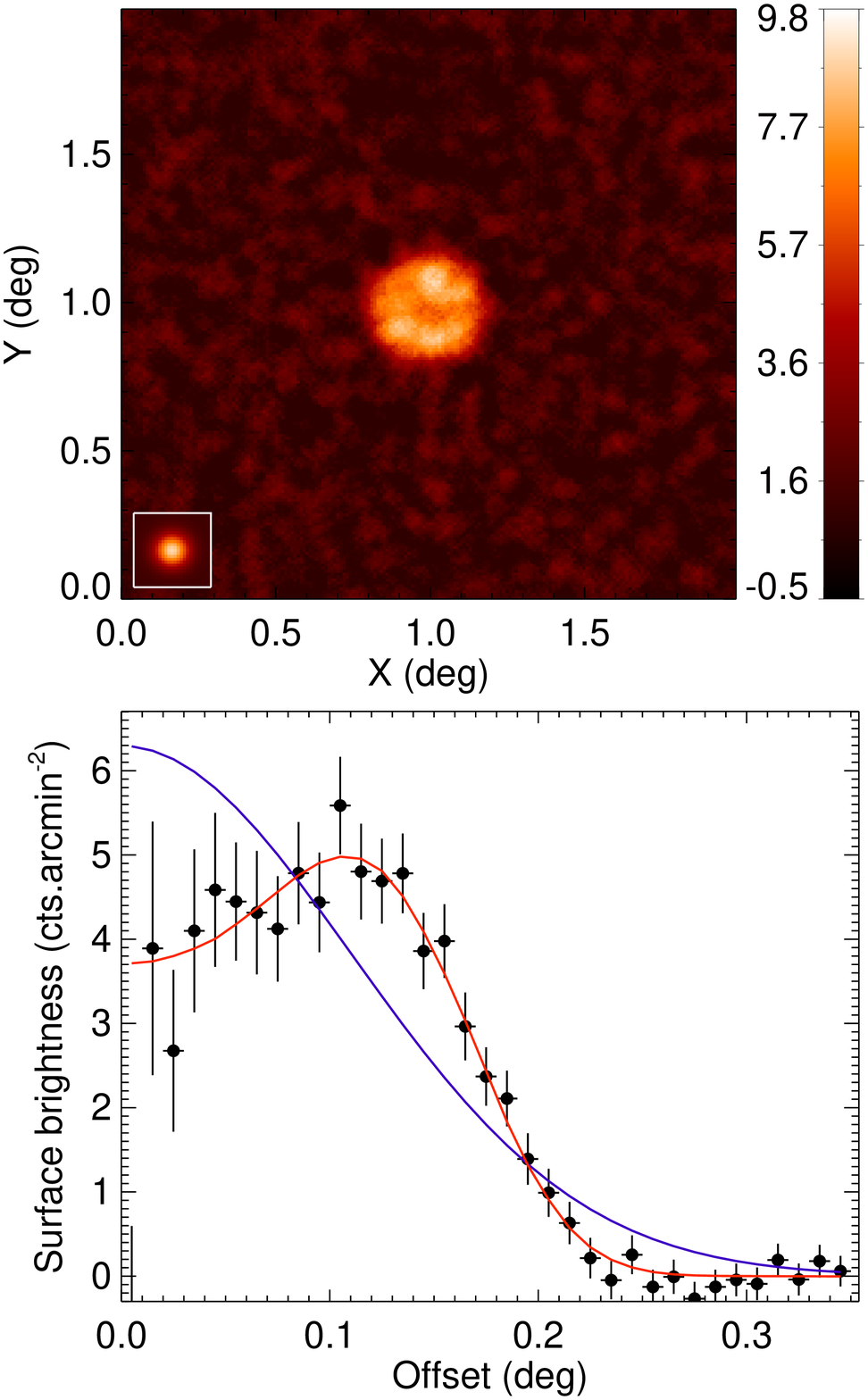}
\caption{\footnotesize {\it Upper panels:} CTA-I simulated images of \rxj-like
         SNR located at 1 (left) and 4 (right) kpc. The color bars give the 
         correlated significance, and the CTA-I PSF is shown in the insets. 
         {\it Lower panels:} Radial profiles of the VHE surface brightness at 
         1 (left) and 4 (right) kpc. The blue and red lines represent the 
         best-fit curves for uniform-sphere and shell-type morphologies, 
         respectively.
}
\label{f:fig3}
\end{figure}

\begin{figure*}[htb!]
\centering
\includegraphics[width=0.66\textwidth]{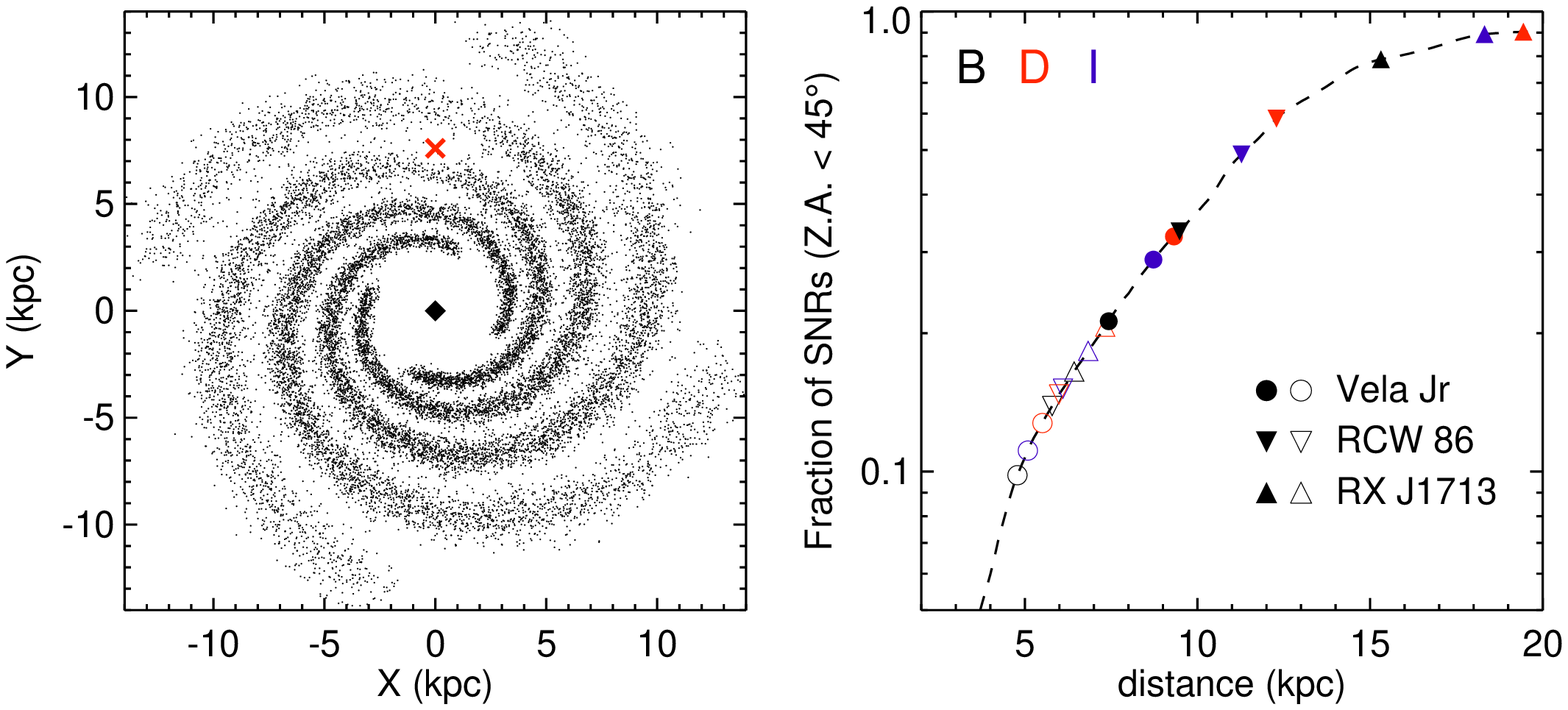}
\includegraphics[width=0.33\textwidth]{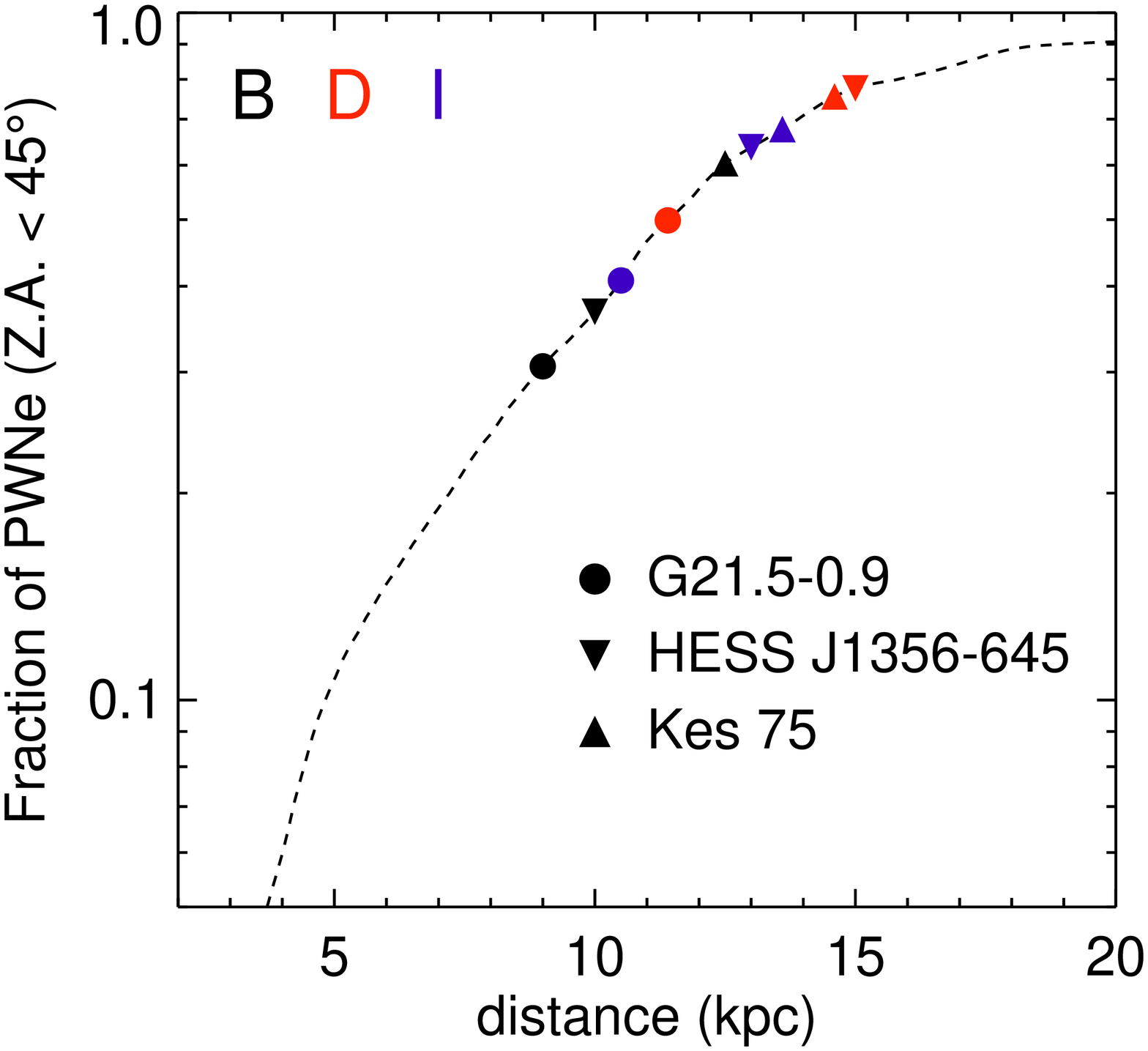}
\caption{\footnotesize {\it Left:} Face-on view of the simulated Galactic 
         distribution of shell-type SNRs and PWNe (see text). The Sun and the 
         Galactic Center are marked with a red cross and a black diamond, 
         respectively. {\it Middle:} Fraction of visible shell-type SNRs 
         as a function of the distance to the Sun. For each source and each 
         CTA configuration considered here, filled and open symbols give 
         respectively the horizons of detectability and resolvability, as 
         defined in the text. {\it Right:} Same as the middle figure, for the
         three considered PWNe. 
}
\label{f:fig4}
\end{figure*}

For this purpose, three shell-type SNRs (\rxj, \velajr, \rcw) and three PWNe
(G21.5$-$0.9, Kes~75, HESS~J1356$-$645) were considered to be representative 
of the SNR/PWN Galactic populations in the VHE domain. Their morphological and 
spectral characteristics, as measured with \hess, together with their 
respective distance estimates, have been used to simulate sources throughout 
the inner ($|\ell|$ $<$ 60\d, $|b|$ $<$ 5\d) Galaxy (see Figure \ref{f:fig3}, 
upper panels). The horizon of {\it detectability} is defined as the distance 
at which the source has a peak significance of 5, while the horizon of 
{\it resolvability} (for shell-type SNRs) represents the distance up to which 
a shell-type fit on the source radial profile (see Figure \ref{f:fig3}, bottom)
is favored at $\geq$ 3 $\sigma$ over a simple gaussian fit (\ie the shell is 
identified as such, and statistically favored over a PWN-like shape). These 
two horizons can be translated into fractions of the total number of 
detectable/resolvable sources with CTA, based on a model of the Galactic 
source distribution. The logarithmic spiral arms model of \citet{c:vallee08} 
is assumed, with a galactocentric distribution given by \citet{c:case98}, and 
an arm dispersion as a function of the galactocentric radius following the 
Galactic dust model of \citet{c:drimmel01}\footnote{With regard to PWNe, any 
potential displacement from pulsar birth place due to the kick velocity has 
been ignored.}. The resulting source distribution is shown in Figure 
\ref{f:fig4}, together with the fraction of visible\footnote{The visibility is 
defined here as the fraction of the sky seen by an observatory located at the 
same latitude as the \hess~site, at zenith angles $\leq$ 45\d.} SNRs and PWNe 
as a function of the distance to the Sun.

As seen in Figure \ref{f:fig4} (middle), a large fraction f$_{{\rm SNR}}$ 
($\sim$0.3--0.9) of RXJ1713-, VelaJr- and RCW86-like Galactic SNRs should be 
{\it detectable} with CTA-I/D (both configurations being definitively favored 
in comparison to B, discarded in the following), but only a small fraction of 
them ($\sim$0.1--0.2) would actually be {\it resolvable}. Although the 
timescale ($\tau_{{\rm SNR}}$) during which a SNR shines in the VHE domain is 
very sensitive to many parameters (related to the stellar progenitor, the 
acceleration process and the ambient medium), it is assumed to be identical 
among all the SNRs. Therefore, the fractions given above are converted into 
numbers following N$_{{\rm SNR}}$ $\sim$ 75~f$_{{\rm SNR}}$~($\tau_{{\rm SNR}}$/3kyr)~($\nu_{{\rm SN}}$/2.5), where $\nu_{{\rm SN}}$ is the Galactic SN rate 
\citep[in SNe per century, see][]{c:li11}. This leads to $\sim$20--70 
detectable TeV SNRs, among which $\sim$7--15 would be resolved with CTA-I/D. 
The same simulations have been carried out by improving the CTA PSF by a 
factor of 2 (\ie $\lesssim$ 1\m~at 1 TeV, see Figure \ref{f:fig1}) in order to
quantify the effect of the angular resolution on the number of detectable 
(N$_{{\rm detect}}$) and resolvable (N$_{{\rm resolv}}$) SNRs. While 
N$_{{\rm detect}}$ gets larger by only 20\% at most, N$_{{\rm resolv}}$ 
increases by a factor of (1.5--1.9).

As for PWNe, a large fraction f$_{{\rm PWN}}$ ($\sim$ 0.4--0.8) of G21.5-, 
HESSJ1356- and Kes75-like nebulae should be detectable with CTA-I/D (see Figure
\ref{f:fig4}, right). \citet{c:dd09} have estimated the lifetime of 
TeV-emitting leptons in such sources to be $\sim$40 kyr (for B = 3 $\mu$G, 
similar to what has been found in several PWNe such as Vela~X). This gives 
N$_{{\rm PWN}}$ $\sim$ 800~f$_{{\rm PWN}}$~($\tau_{{\rm PWN}}$/40kyr)~($\nu_{{\rm PSR}}$/2), and implies that $\sim$ (300--600) PWNe should be detected with 
CTA-I/D. Again, these estimates should be taken with care as they do not 
account for any time evolution of the TeV luminosity in these sources, which 
itself depends on many parameters \citep[\eg][]{c:gelfand09}.

\section{Conclusions}
 
In this contribution, simulations of Galactic shell-type SNRs and PWNe in the 
VHE domain as seen with CTA have been presented, for three different array 
configurations. As shown above, the configurations optimized at very-high 
energies ($>$ 1 TeV) are clearly favored with respect to these two source 
classes. This is naturally the case for probing the nature of the 
highest-energy gamma-rays in the brightest SNRs such as \rxj~(and more 
generally for seeking in the $\gtrsim$ 100 TeV domain the so-called Galactic 
PeVatrons, \ie sources of CRs with energies close to the knee at $\sim$ 3 PeV).
Moreover, first estimates of the number of detectable/resolvable SNRs and PWNe 
have pointed out the twofold importance of the angular resolution, in order to 
(1) pinpoint the shell-type morphology of SNRs and (2) mitigate the source 
confusion in the inner regions of the Galactic Plane, given that hundreds of 
PWNe are expected to be revealed with CTA. For this purpose, an angular 
resolution of $\lesssim$ 1\m~at TeV energies seems to be desirable, although
more quantitative estimates of the source confusion need to be carried out. 
As for the former point, it turns out that, regarless of the final instrument
PSF, a large fraction of the detected shell-type SNRs will not be identified 
as such based on solely their VHE morphology. Therefore, follow-up 
multi-wavelength observations with the on-going or planned facilities (EVLA, 
LOFAR, eROSITA, NuSTAR, ASTRO-H), towards these potential CR sources will be 
of crucial importance.

\begin{acknowledgements}
I thank all the members of the CTA-PHYS working group for valuable discussion
and helpful comments on the manuscript. We gratefully acknowledge support from 
the agencies and organisations listed in this page: \\
http://www.cta-observatory.org/?q=node/22.
\end{acknowledgements}

\footnotesize

\bibliographystyle{aa}

\end{document}